\documentclass[]{article}
\usepackage{lmodern}
\usepackage{amssymb,amsmath}
\usepackage{ifxetex,ifluatex}
\usepackage{fixltx2e} 
\ifnum 0\ifxetex 1\fi\ifluatex 1\fi=0 
  \usepackage[T1]{fontenc}
  \usepackage[utf8]{inputenc}
\else 
  \ifxetex
    \usepackage{mathspec}
  \else
    \usepackage{fontspec}
  \fi
  \defaultfontfeatures{Ligatures=TeX,Scale=MatchLowercase}
\fi
\IfFileExists{upquote.sty}{\usepackage{upquote}}{}
\IfFileExists{microtype.sty}{%
\usepackage{microtype}
\UseMicrotypeSet[protrusion]{basicmath} 
}{}
\usepackage[margin=1in]{geometry}
\usepackage{hyperref}
\hypersetup{unicode=true,
            pdftitle={Temporal evolution of notoriety of Wikipedia pages with origin in social networks},
            pdfauthor={JJ Merelo-Guervós and Elena Merelo-Molina},
            pdfborder={0 0 0},
            breaklinks=true}
\usepackage{graphicx,grffile}
\makeatletter
\def\maxwidth{\ifdim\Gin@nat@width>\linewidth\linewidth\else\Gin@nat@width\fi}
\def\maxheight{\ifdim\Gin@nat@height>\textheight\textheight\else\Gin@nat@height\fi}
\makeatother
\setkeys{Gin}{width=\maxwidth,height=\maxheight,keepaspectratio}
\IfFileExists{parskip.sty}{%
\usepackage{parskip}
}{
\setlength{\parindent}{0pt}
\setlength{\parskip}{6pt plus 2pt minus 1pt}
}
\ifx\paragraph\undefined\else
\let\oldparagraph\paragraph
\renewcommand{\paragraph}[1]{\oldparagraph{#1}\mbox{}}
\fi
\ifx\subparagraph\undefined\else
\let\oldsubparagraph\subparagraph
\renewcommand{\subparagraph}[1]{\oldsubparagraph{#1}\mbox{}}
\fi

\let\rmarkdownfootnote\footnote%
\def\footnote{\protect\rmarkdownfootnote}

\usepackage{titling}

  \title{Temporal evolution of notoriety of Wikipedia pages with origin in social
networks}
  \pretitle{\vspace{\droptitle}\centering\huge}
  \posttitle{\par}
  \author{JJ Merelo-Guervós and Elena Merelo-Molina}
  \preauthor{\centering\large\emph}
  \postauthor{\par}
  \predate{\centering\large\emph}
  \postdate{\par}
  \date{March 9th,2017}

\begin{document}
\maketitle
\begin{abstract}
The Wikipedia is a web portal created by users and its simplicity,
references and also the inclusion as insets introductory paragraphs for
their pages in Google search results have made it the go-to place to
find out about current events or people featured in them. Besides, its
open application programming interface (API) allows any user to know
about the number of visits some particular page has. In this paper,
after certain events that made Copernicus a viral meme in Spain, we
study the intensity and duration of the increment of visits to his page
and other pages related to the event. Using pages related to other
persons as a comparison, we try to establish a typical duration of
notoriety achieved through social networks, mainly comparing with visits
in previous years in the same dates. We conclude that a 7 day duration
is the statistical mode and that this duration is relatively independent
of the initial increase in the number of visits.
\end{abstract}

\section{Introduction}\label{introduction}

The so called ``Slashdot Effect'' (Adler 1999,Halavais (2001)), known
since the end of the 90s, is the name attributed to the increase in
pageviews that the websites experience when mentioned by publications
like \href{http://slashdot.org}{Slashdot}, one of the most visited in
the area of technology and especially programming. This effect provokes
a peak in visits, especially directed to a single webpage, that in a lot
of cases and more obscure ages in which there were not any publication
of content networks, had a similar effect of a denial of service attack.
In Spain we have also referred to the ``Efecto Menéame'' (Dans 2008),
because of the influence of this aggregator of news and even, before and
in a lesser measure, the ``Efecto Barrapunto'' (Blanco 2009) originated
by mentions to the web Barrapunto.com, a Spanish Slashdot.

However, to the extent of our knowledge, until now there has not been
many studies analyzing a similar effect aroused by mentions in social
networks, maybe due to the fact that such an effect is not caused by a
single website and might be eventually directed to many different pages;
even lesser is the effect they have in the search or mindshare of the
concepts. Indeed, web analytics is currently far more complex and
generally the visits on a page, say, of a newspaper or ecommerce site
are not public. And besides, instead of studying the effect of a single
link, nowadays it is of more interest to study the topics that pop up in
the social networks. Nevertheless, the effect these have in the
mindshare of the public and its behaviour while surfing the Net has not
received so much attention as the behavior of the trending topics
themselves.

At the start of 2017, an anecdote in a Wise Men from Orient parade in
Madrid, in which a video newscaster referred to a character in a float
as Colon, to which he answered ``I am Copernico!'' in an amusing tone,
\href{http://verne.elpais.com/verne/2017/01/07/articulo/1483785572_720611.html}{as
is narrated (in Spanish) in Verne}, gave way to a storm of memes, which
increased on the following days when a journalist and a politician
discussed in a TV program about what did Copernico really do.

In a series of articles and figures written in Spanish (Merelo 2017b,
Merelo (2017c), Merelo (2017a),Benito (2017)) that show the temporary
evolution of this mindshare through the visits to the Wikipedia page of
Copernico (Wikipedia 2017) we have analyzed the change in the number of
visits, especially looking for the duration of this phenomenon by means
of studying the daily number of visits. In this article, that closes the
series, we try to find the causes of that duration, comparing it with
other phenomenons of visits to the Wikipedia which have happened in a
similar time frame.

As far as we know, there have been no studies that focus on the temporal
dimension of an Internet meme. (Adamic et al. 2016) focuses on how
information evolves and how this evolution is time dependent, and
(Naaman, Becker, and Gravano 2011) analyzes other features such as the
decay in the number of occurrences of trending topics and how people
interact with it.

The rest of the paper is organized in the following way: first we
present the methodology that has been followed in order to capture and
study the data is presented. In the next section the visits to the
Wikipedia page of Copernico and other related webpages is analyzed,
along with the evolution in its ranking of viewed pages in the
Wikipedia. We will also examine the visits to the pages of other persons
that have become notorious due to several reasons. Finally, we will draw
some conclusions.

\subsection{Methodology}\label{methodology}

The Wikipedia, through its application programming interface (API),
allows access to the visits of each and every one of the different pages
it hosts. It is a REST interface which can be directly accessed without
any authentication. For the purposes of this paper, we tap this API
using a script that, along with the rest of data and programs, is free
software and hence can be downloaded from GitHub (Merelo 2017a). The
script uses \texttt{curl} for downloading the file, just by codifying as
URL the name of the page, next to \texttt{all-agents}, and selecting as
parameters the two dates in which we are interested, from the start of
2017 until now. Due to the fact that the Wikipedia API returns a file in
JSON, we pull the data we are interested in out by means of the
\texttt{jq} utility, that allows to make complex queries to JSON files.
That way, we extract the views, and we put it in data files, which will
also be attached to this paper. The rest of the data analyzed in this
paper is extracted in the same way, via shell scripts that download an
URL and the use of \texttt{jq}.

Nonetheless, in order to extract the ranking of the one thousand most
visited pages, a script that gets the position in the ranking and the
number of visits, along with the title of the article, is used. Then,
the information is filtered, leaving only the ``content'' articles that
are in the ranking, deleting special pages. Furthermore the ranking is
once again calculated. Last, these rankings are processed to calculate
the evolution in the position of the page as the days go by, saving in a
file a number with the position that a specific page occupies each day.

The results of analyzing the data are shown in the following section.

\subsection{Visits to the page of Nicolás Copérnico and other related
pages}\label{visits-to-the-page-of-nicolas-copernico-and-other-related-pages}

Data is processed in this paper using R; scripts can be accessed in the
source to this paper.

First we will focus on the anecdote that originated this paper by
analyzing the visits to the Nicolás Copérnico page in the Spanish
Wikipedia during 2017 and the same period in 2016. This period falls
outside the school year in Spain, with classes starting just next to
that day. That might account for a certain interest in the page, which
implies that there will be some daily variation in the number of visits
anyway. Comparing the visits for two years will help us discount that
effect.

\includegraphics{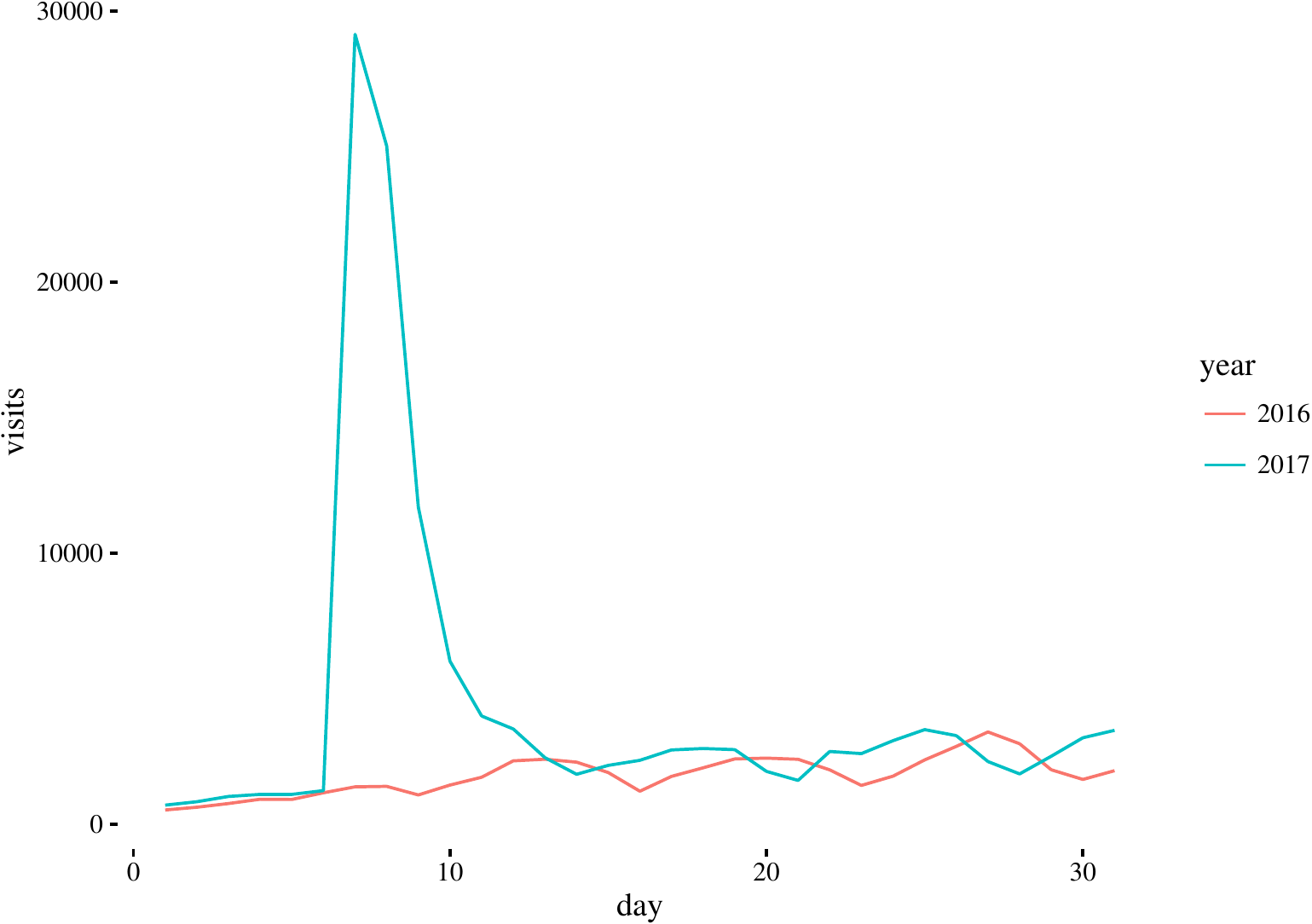} It is
noticeable that, after reaching a peak the first day, it descends the
second one and even quicklier the third one. Nevertheless, from the
fourth day the fall is much softer, until having approximately the same
data a week later. In fact, there is some variation also in the previous
year, that is why we compare the relationship between last and this
year´s visits in the next chart.

\includegraphics{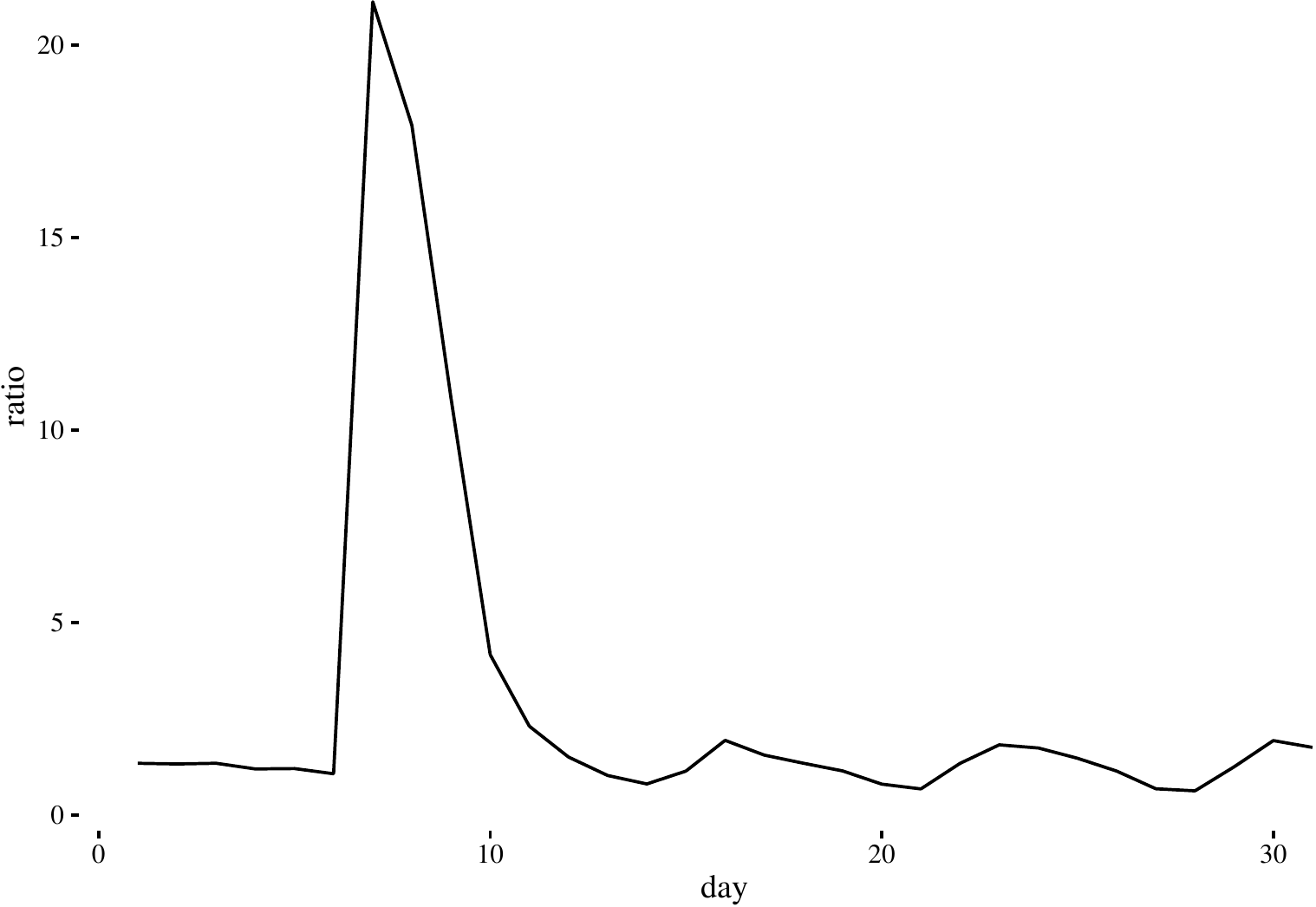} From an
initial situation where the number of visits is similar, though slightly
superior, it goes to having 20 times more visits, number that slowly
drops to 10 times three days past the peak. On the fifth day the rate is
of a 50\% more visits; a little bit over last year´s on the same date,
taking into account that last year´s tendency was of a gradual growth in
the number of visits. Eventually, the number of visits reaches a
plateau.

It is interesting to have a look at the evolution of some pages linked
to the same concept, especially the one of the main achievement of
Copérnico, the heliocentric theory, which motivated the following day a
debate on TV.

\includegraphics{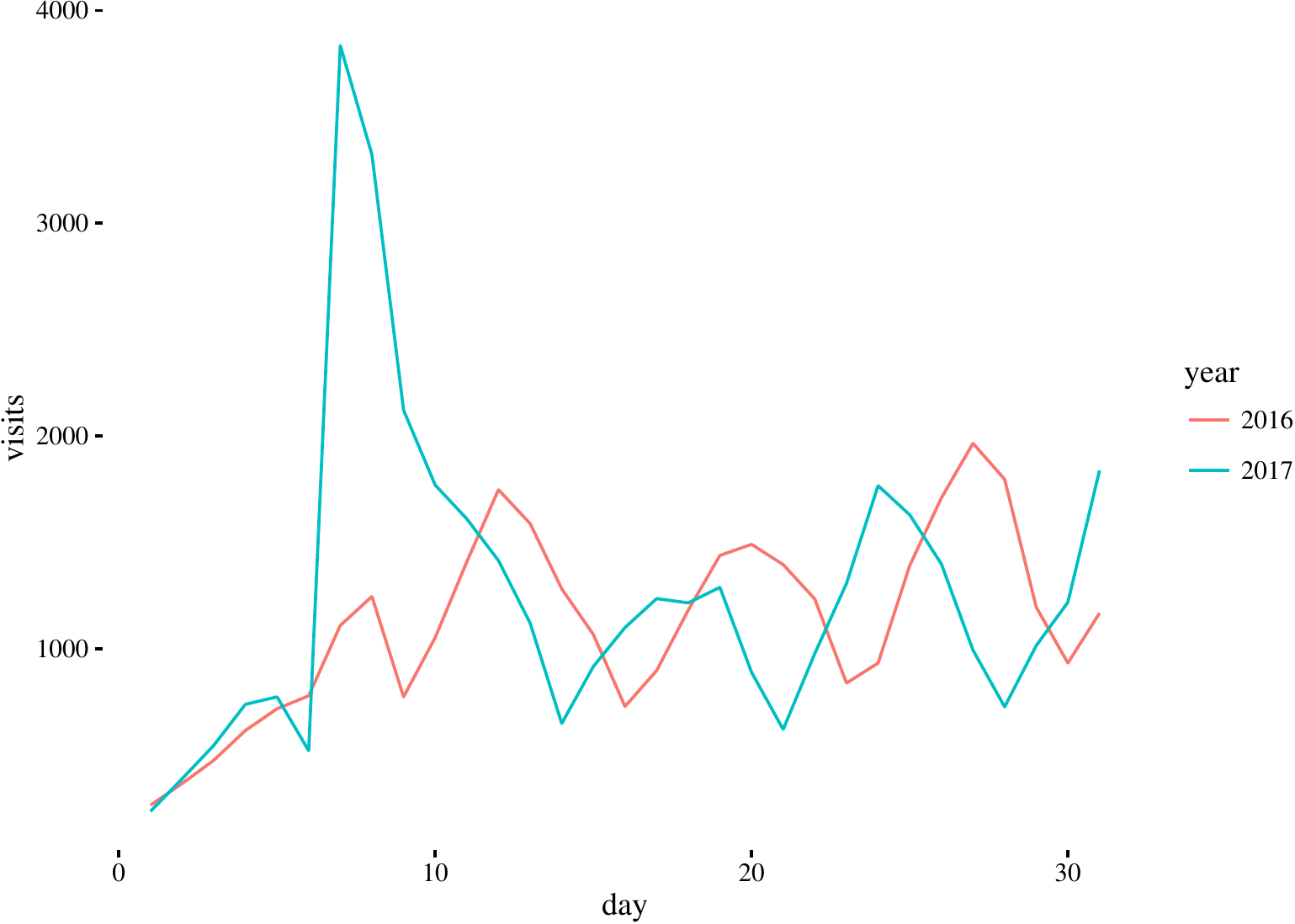} The number
of visits moves in a different scale and thus there are more
fluctuations along the day. However, like in the previous case, even
though initially the visits were very much alike or even inferiors,
presumably in five days it went back to its initial value, with a small
variation. In five days it has less visits than last year, the same
situation that the previous day to the incident.

What also changes is the correlation between the number of people that
visit a web and the ones that visit the other. We have a look at it in
the next graphic.

\includegraphics{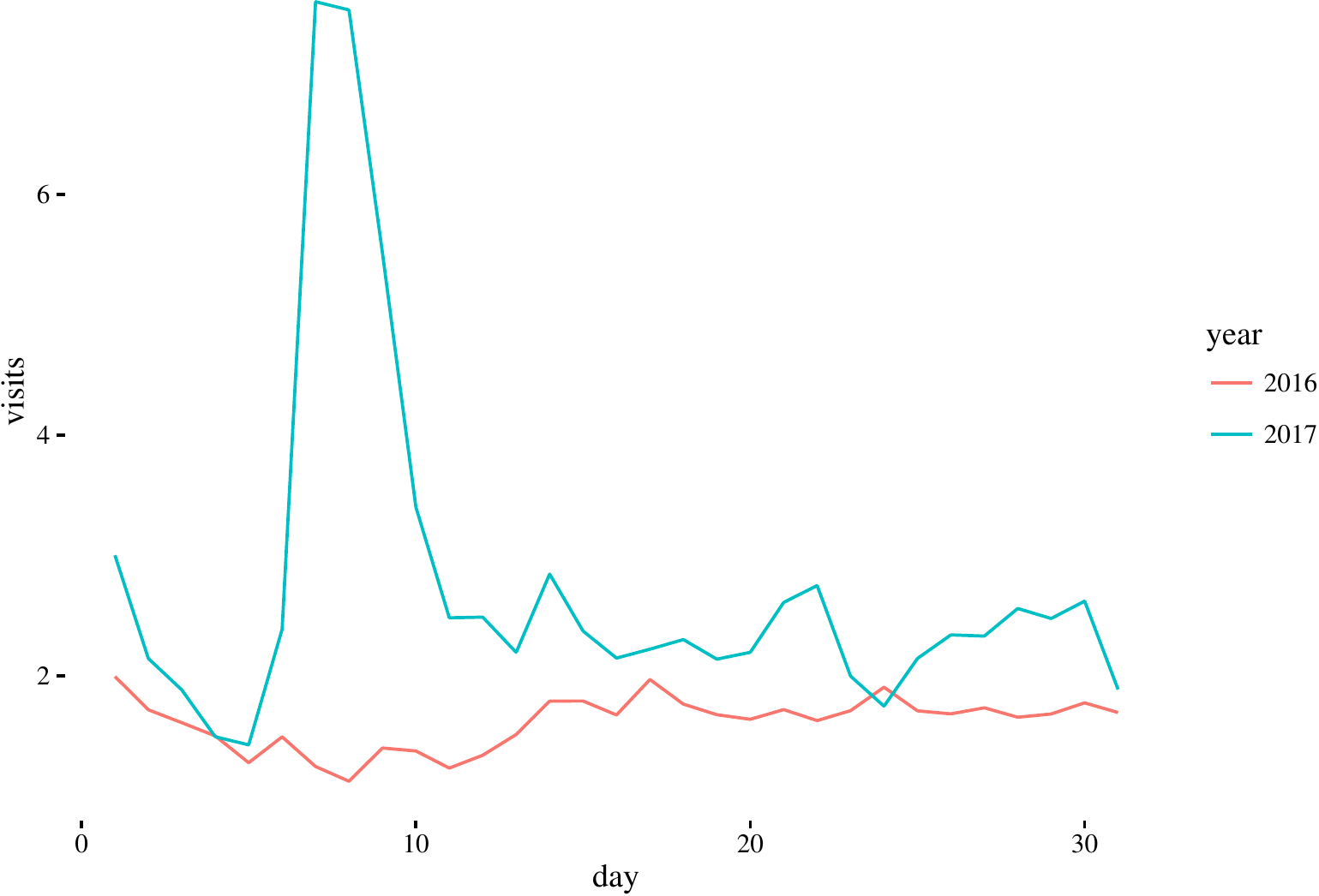}

In the year 2016 there is a rate inferior to 2, consequently the page of
the heliocentric theory receives more or less half the visits that
Copérnico´s, with little variations. We suppose that there is more
interest in the character than in its contribution, though not a lot. On
the other hand, when the issue ``Copérnico-Ojeda'' breaks out, all of a
sudden the people get curious about the character, without caring much
about his contribution to the scientific world. It is important to point
out that we do not know whether they are the same people that visit the
webpage, because of the fact that the path they follow in Wikipedia is
not returned by the API. Anyway, it is reasonable to suppose that a big
percentage of the people that visit whichever of the pages will end up
following the links in order to visit the other one. The link to the
heliocentric theory appears in the first lines of Copérnico´s page and
it is the thirteenth link from the start of the page, while Copérnico
shows up int the second paragraph of the theory page, being the tenth
link in this case.

Let´s also see another page, Galileo Galilei´s, who spread the
heliocentric theory. His name does not show up in any way in the
incident but checking the visits here would be relevant to study the
interest these kind of pages have and whether there is a offshoot of
interest from one to another.

\includegraphics{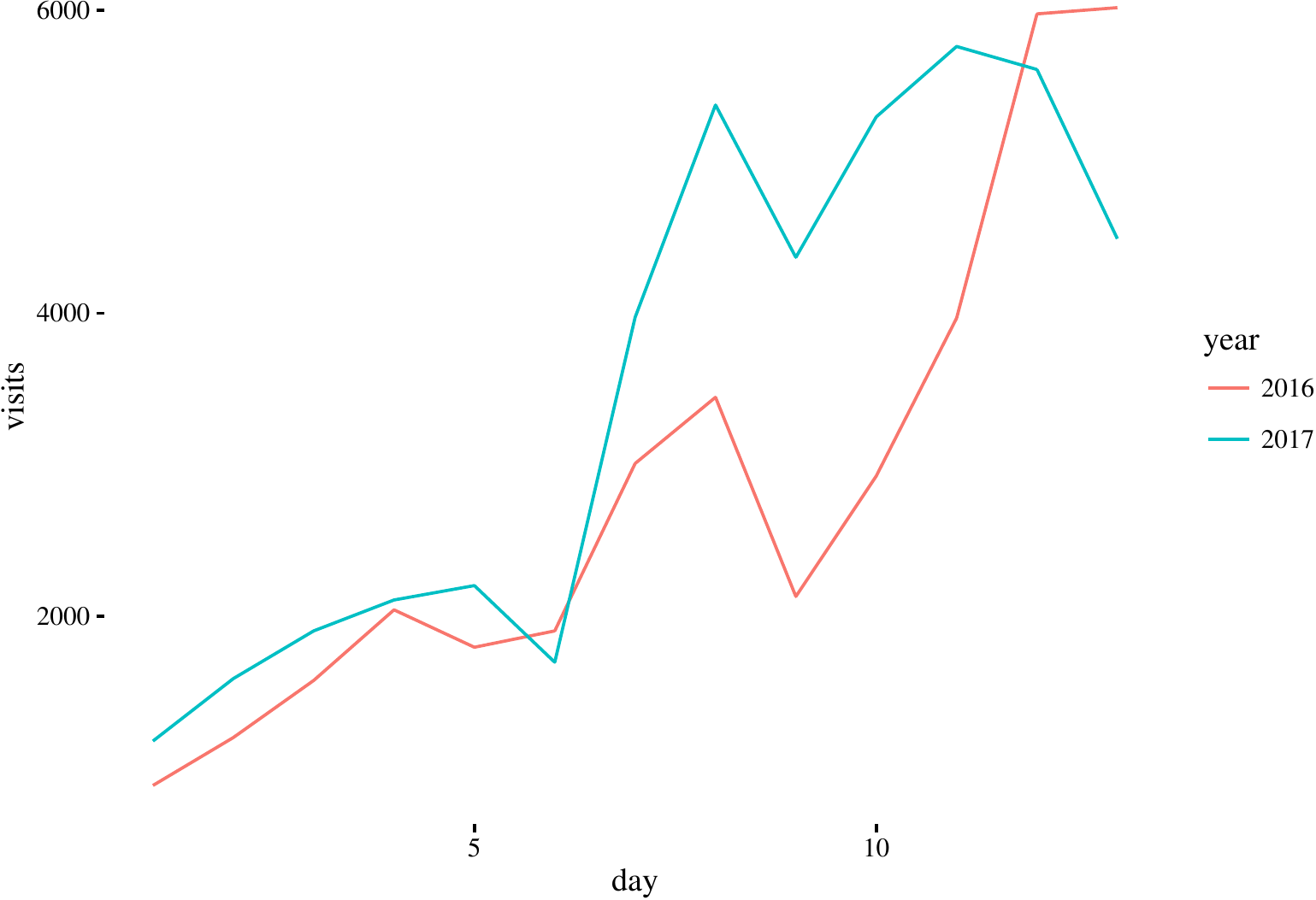}

Apart from the curiosity on how do they evolve in the same way, with a
considerable drop the seventh day, there is neither an obvious tendency
nor can we suppose that pages of this kind of content could have an
increased interest. In fact, the twelfth day there were less visits in
Galileo´s page than last year. Although there is a chance of a transfer
of attention and hence of visits from one to another, since Galileo´s is
linked to Copérnico´s, and as there are not changes in the same way as
in Copérnico´s a sudden interest in the Renaissance Science and the
heliocentric theory like a cause of the rise in visits that will be then
attributable to the incident is ruled out.

From this analysis we can conclude that the interest in a particular
Wikipedia page famous through a meme in the social networks lasts around
seven days, and that it is a shallow interest extending exclusively to
the page itself, not any concepts related to it. Let us see next what
happens to other similar pages.

\subsection{Duration of notability in other Wikipedia
pages.}\label{duration-of-notability-in-other-wikipedia-pages.}

We also examine the visits to the page of Meryl Streep, whose speech
during the Golden Globes had a considerable impact, in the same way as
the elected president´s reaction did. In this case it is an example of a
page that on its own has a lot of visits. We plot it side by side the
visits to the page of John Hurt, who died on January 21st, 2017.

\includegraphics{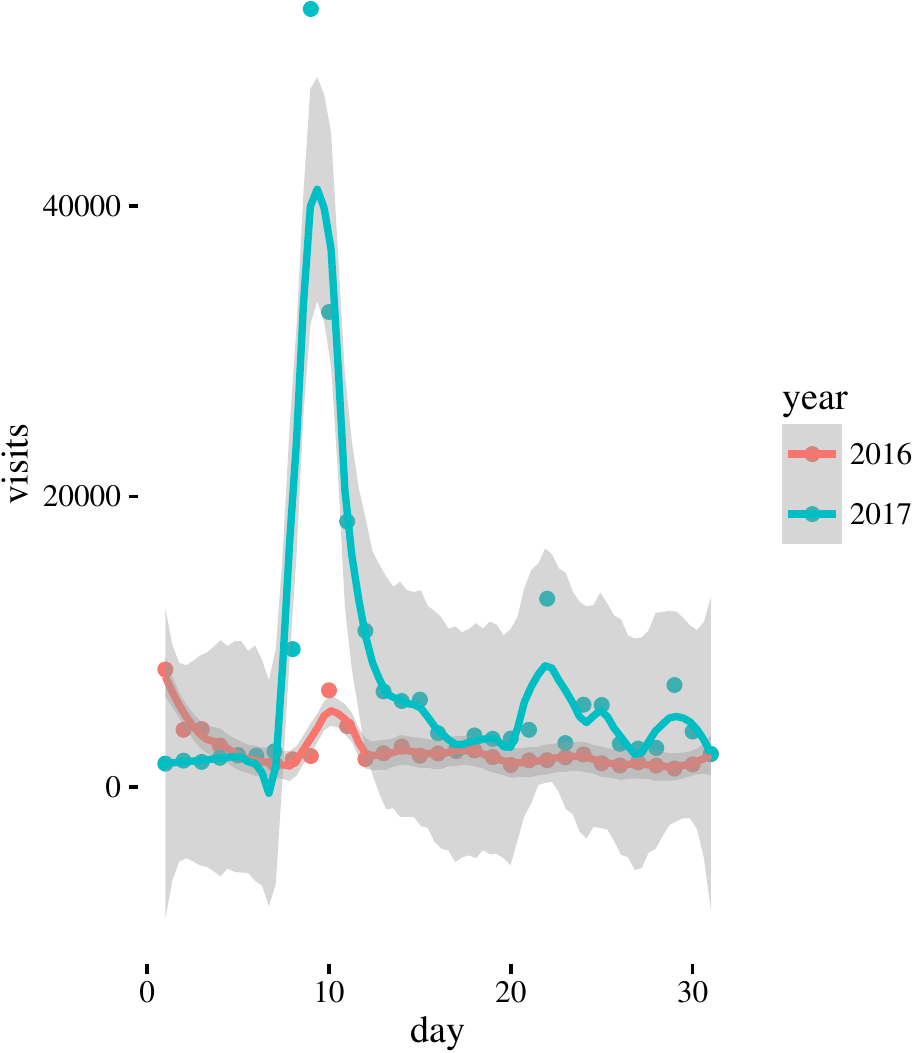}
\includegraphics{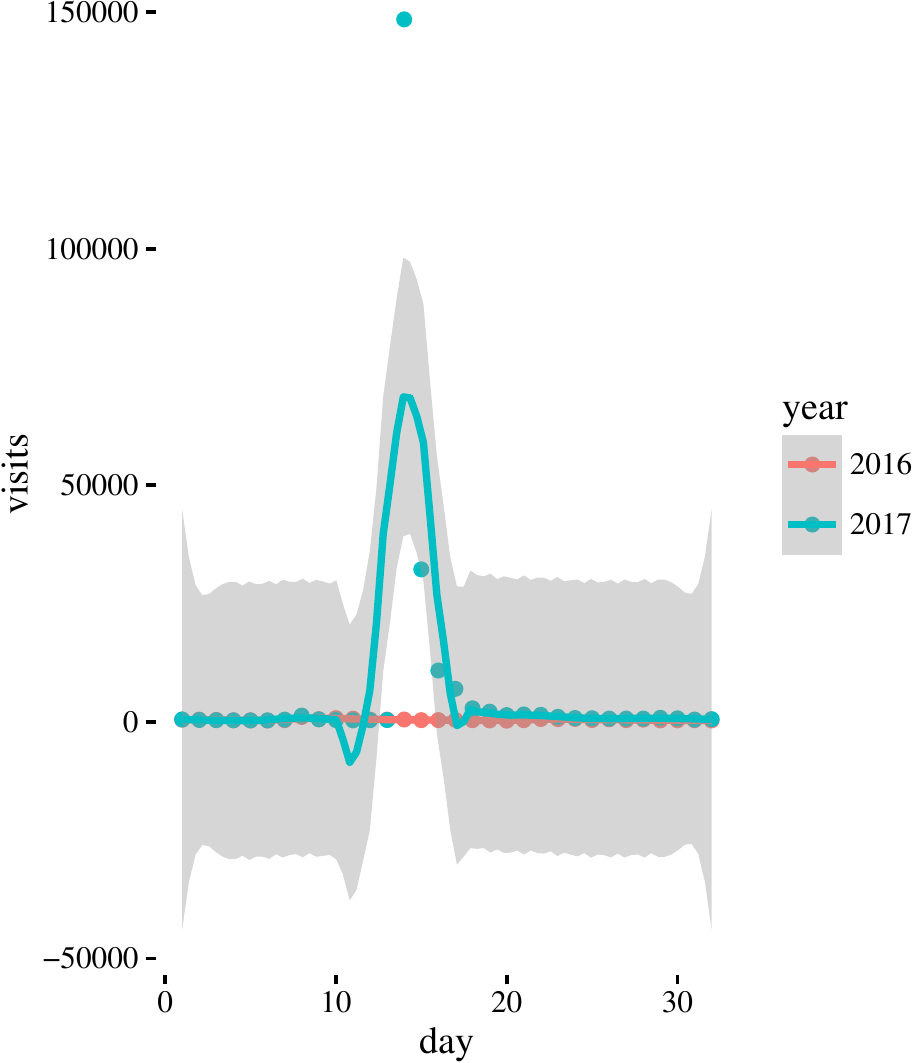}

The number of visits reached by John Hurt's Wikipedia page is almost 3
times as high as the one reached by Meryl Streep. The mechanism at work
here must go in the opposite direction of fame: John Hurt was not
probably known by name, so people looked him up much more than Meryl
Streep, who is a household name and whose movies are shown continuously,
even in Spanish TV channels. But even with the different scales, the
\emph{relaxation time} hovers around 7 days, a few more in the case of
Meryl Streep (left). Besides, when the visits plateau the \emph{new
normal} is slightly superior to the visits achieved in the previous year
or before the incident that initiated the sudden fame. Let us see this
in the second case, the John Hurt page.

\includegraphics{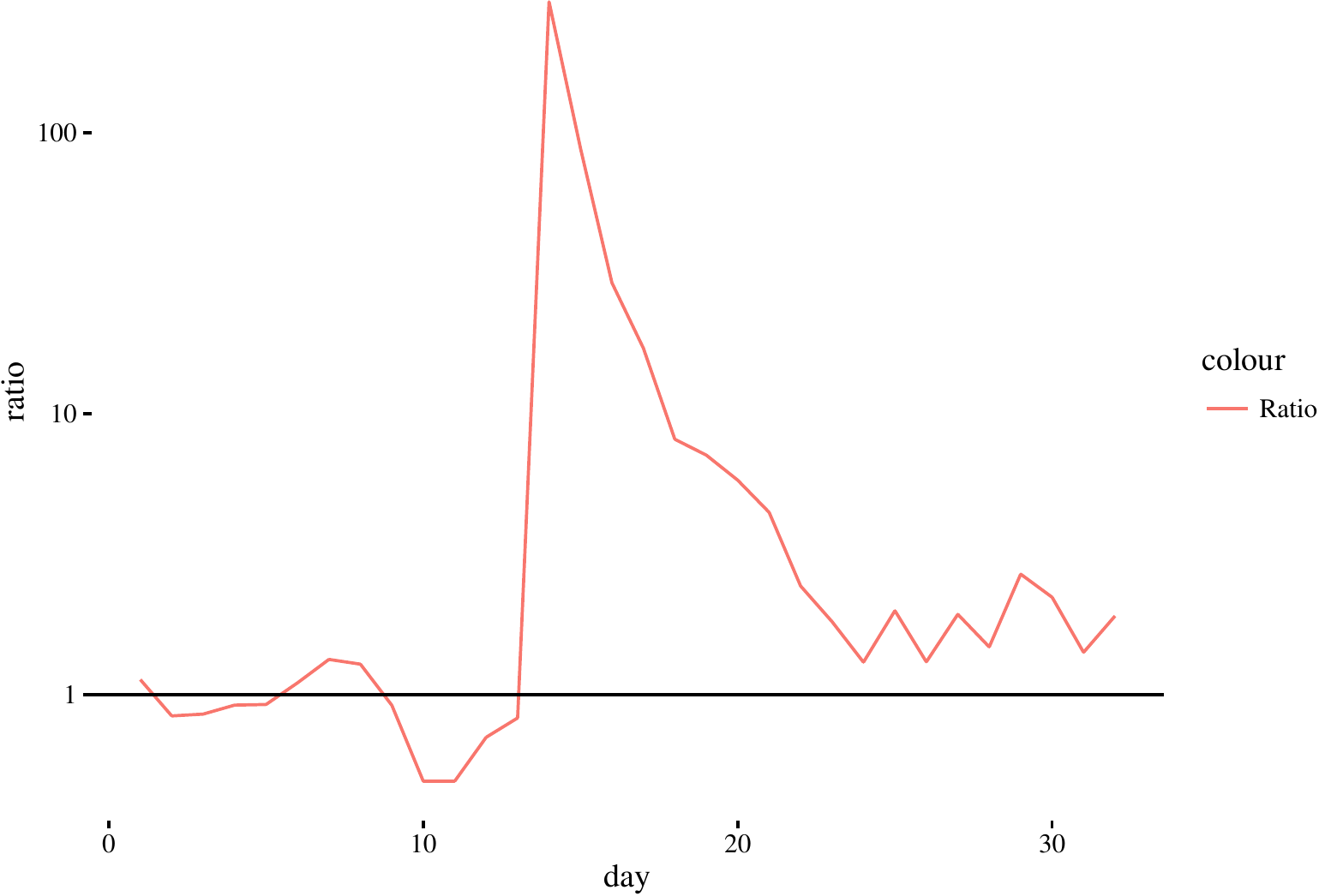}

We plot the \emph{y} axis in a logarithmic scale to see more clearly the
ration after \emph{relaxation}. The straight line goes through the ratio
== 1. Although there are variations, for a few days after the peak ratio
is still slightly higher than 1, which seems to imply that the
\emph{mindshare} of this particular meme follows two different phases:
one of sharp decline, that lasts the aforementioned 7-10 days, followed
by a plateau where the decline is much slower, and stays stationary with
a ratio higher than one, in the same way as was seen before.

These tho regimes, duration and notability need not take the same time,
and will probably vary along time. We have also examined the cases of
three other memes: from left to right, the death of Robert Vaughn in
November 2016 and Carrie Fisher and George Michael in December 2016. Our
intention was, looking at these events taking place a few months before,
to see whether there is any variation.

\includegraphics{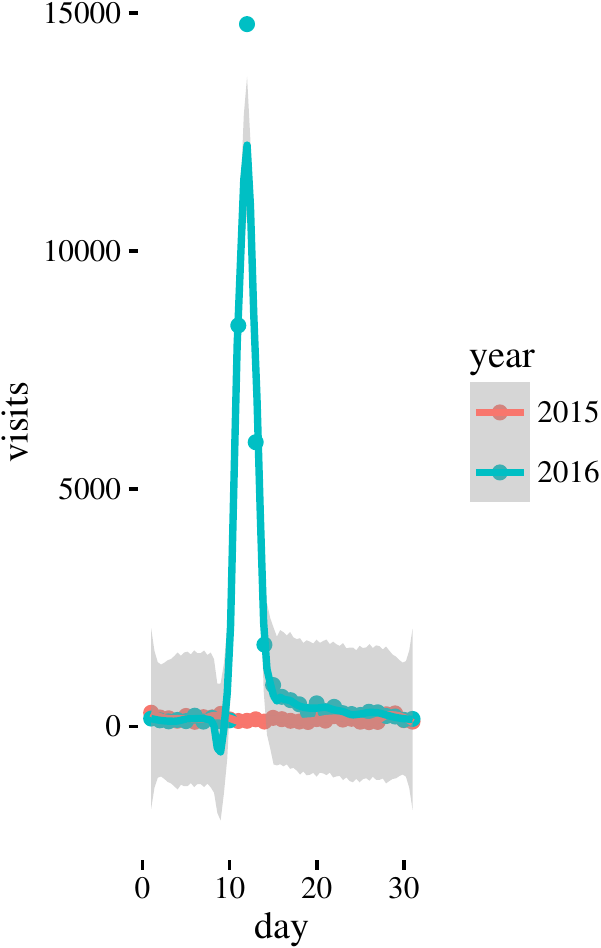}
\includegraphics{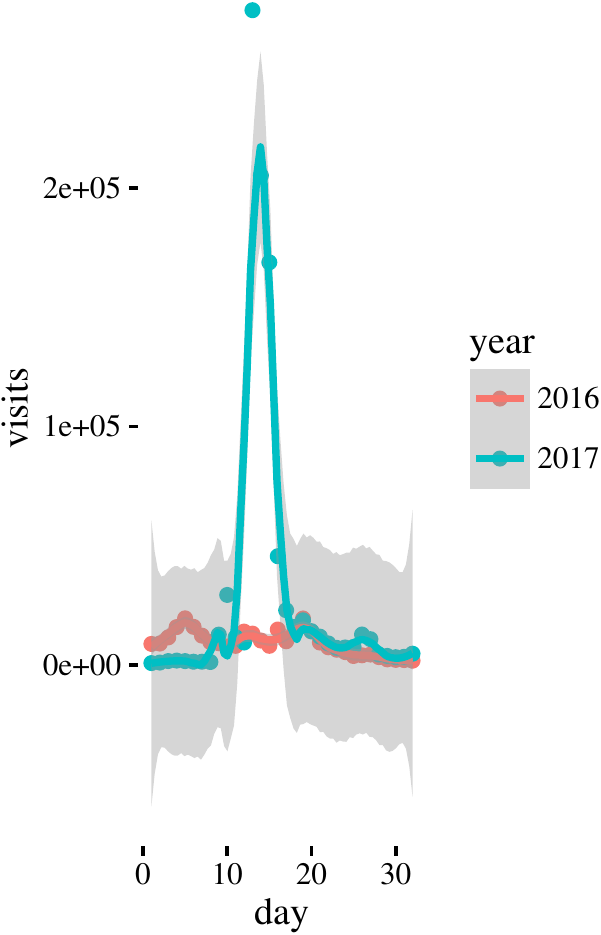}
\includegraphics{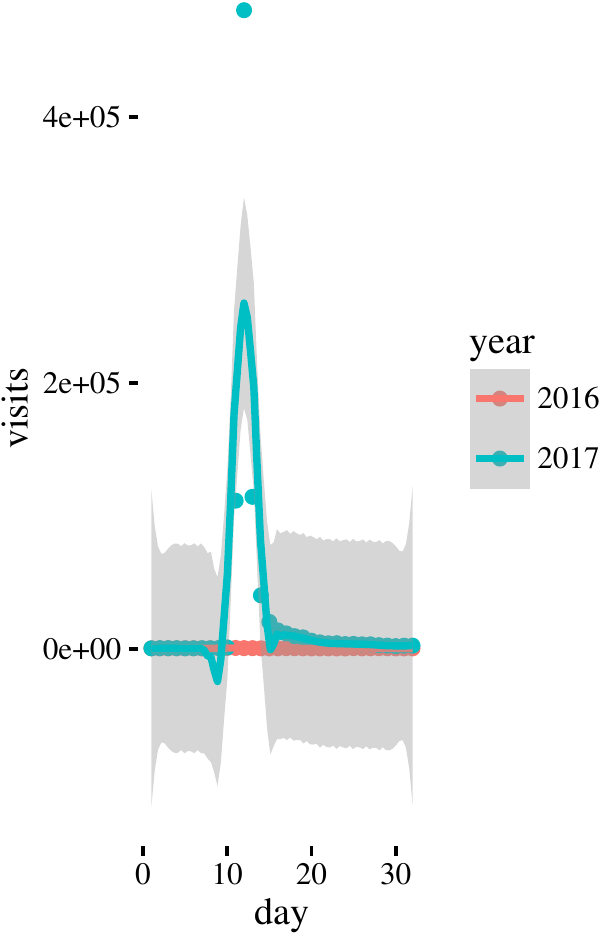}

Let us first look at the first chart on the left, representing visits of
the Robert Vaughn page. Being not so well known in pop culture, the peak
reached is relatively small, but still visits multiplied 100-fold when
his death was known. Carrie Fisher (center) and George Michael (right),
being better known, had a higher level of visits, but the visits reached
multiply 200-fold, and in this case the two regimes can be observed:
sharp decrease that lasts 10 days followed by a plateau where the ratio
to visits before the event is around 2-3. The same type of behaviours
can be observed about the visits to the page of George Michael.

Finally, let us look at similar events that happened later along the
time line: the viral sensation of Winona Ryder showing a range of facial
expressions during a gala (left), and the death of a journalist and
famous 80s musical TV anchor in Spain, Paloma Chamorro (right).

\includegraphics{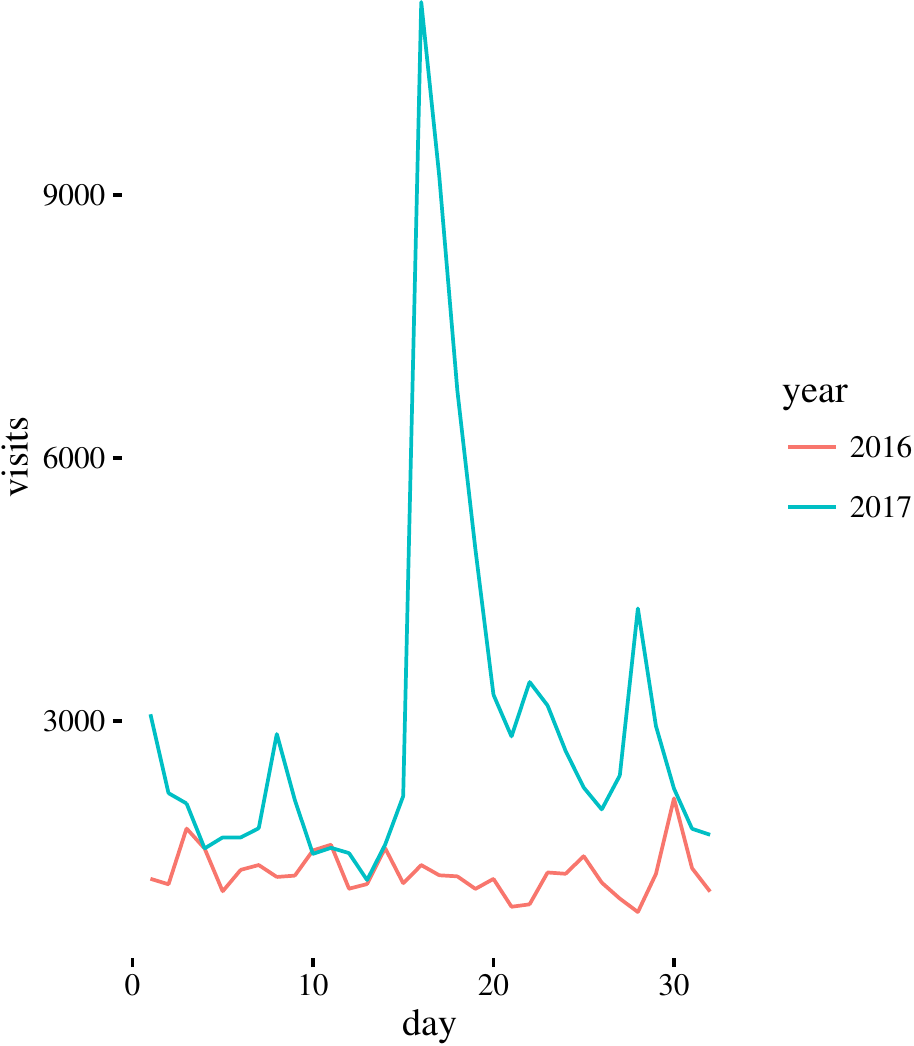}
\includegraphics{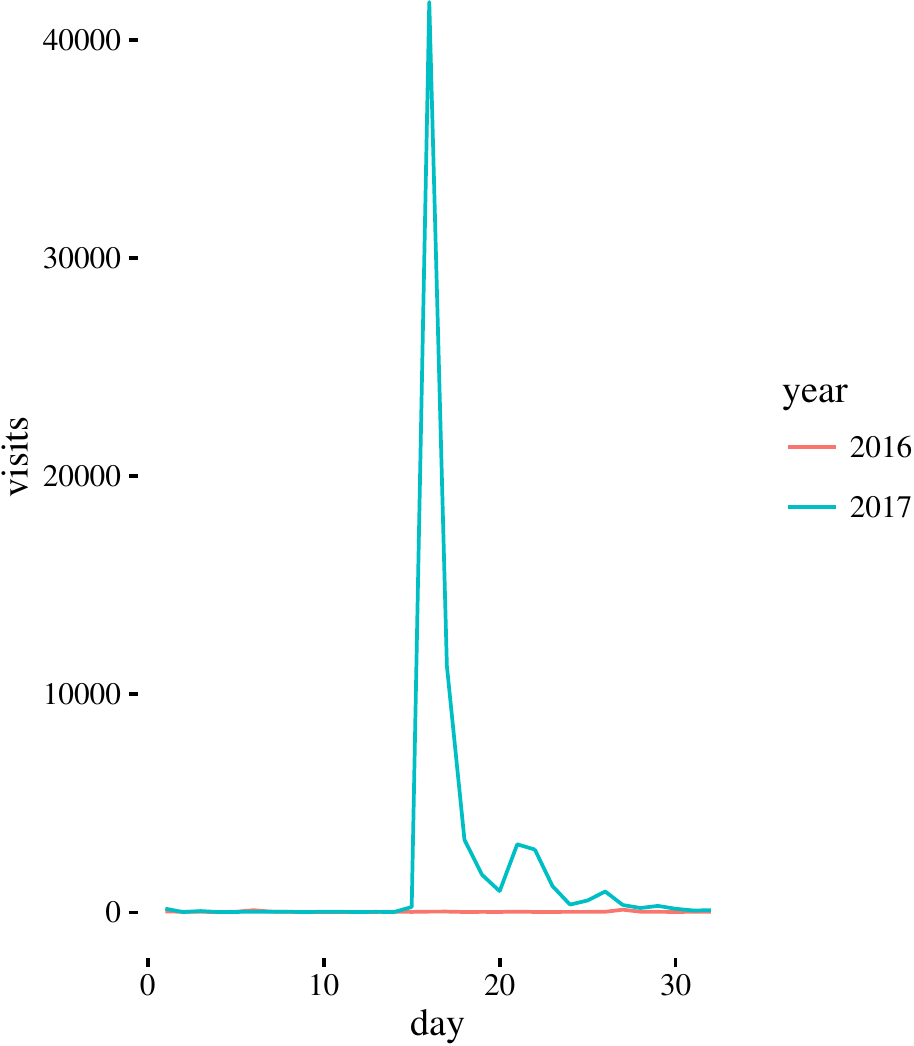} The chart on
the left, although with some \emph{noise}, shows more or less the same
phenomenon of a sharp peak followed by rapid decrease in visits, that
take between 7 and 10 days. The lower number of visits reached probably
accounts for the fact that the meme was self-contained (a series of
faces) and did not need additional information on the person. However,
the one on the right is entirely familiar in scale and size, with
\emph{rebounds} probably due to published obituaries and settling to a
number of visits that is slightly superior than originally.

Eventually, all visit graphs resemble a ``witch´s hat'', with a
noteworthy peak and a more gradual decrease in the number of visits in
approximately a week´s time, although usually a bit longer. We will
examine this findings in the next section, where we present our
conclusions.

\subsection{Conclusion and
discussion.}\label{conclusion-and-discussion.}

By examining the visits of Wikipedia pages through its open API we
intend to look for patterns in these visits and, through them, establish
the duration of Internet memes, assuming that when one particular person
or concept becomes noticeable, people will check out their Wikipedia
page as first source of information.

We have done it by looking at nine different pages corresponding to 9
different persons that have become trending topics in social networks,
though specially in Spain, starting with a very local event that
actually initiated this research. What we have found analyzing visit´s
data is that noteworthiness lasts for at least a week, and sometimes a
bit more, up to 10 days; from then on, visits go back to the usual rate
but at a much slower pace. The headline would thus be ``Fame in social
networks lasts for only a week'', but this would have some caveats that
would be worth checking out in the future, such as the relationship with
the nature of the meme itself (due to an unexpected event in the life of
the person, for instance, or some other thing). A secondary finding is
that the magnitude of the increase in visits is related to prior fame:
``those who have, will get more'', reaching a much higher ration between
prior and post-event visits.

The fact that the duration for this type of memes or trending topics is
in a very narrow band is consistent with the findings of Adamic and
coauthors (Adamic et al. 2016), in the sense that memes evolve and its
original intention or reference concepts change in time; they are also
obviously substituted by other, more recent, events. The fast initial
decay and lower decrease afterwards is also consistent with the study of
(Naaman, Becker, and Gravano 2011) over a wide range of trending topics,
although no measurements on this duration are made. In this sense, this
is the first study, although focused on a few and mostly regional
events, that has studied the duration and mid-term secondary effects of
memes on society's mindshare as reflected in visits to Wikipedia pages.

In general, the use of this API by Wikipedia opens a whole world of
possibilities for the research of collective behavior. One of them would
be to analyze what part of these visits were done using different
devices, and which ones correspond to edits or simple visits. The
difference between the rise in visits of life-changing events or other
kind of memes might be due to this fact. This could be an interesting
venue of research in the near future, as well as checking if there is in
fact some change in this one-week duration of memes.

\subsection{Note}\label{note}

All files, data and scripts needed to generate this paper are available
at \href{http://github.com/JJ/Cop-rnico-visitas}{the GitHub repository
for this paper} and can be used with a free license. If you use in any
scientific publication we are grateful for referencing this paper or the
other papers working on the same data included in the bibliography.

\subsection*{References}\label{references}
\addcontentsline{toc}{subsection}{References}

\hypertarget{refs}{}
\hypertarget{ref-adamic2016information}{}
Adamic, Lada A, Thomas M Lento, Eytan Adar, and Pauline C Ng. 2016.
``Information Evolution in Social Networks.'' In \emph{Proceedings of
the Ninth Acm International Conference on Web Search and Data Mining},
473--82. ACM.

\hypertarget{ref-adler1999slashdot}{}
Adler, Stephen. 1999. ``The Slashdot Effect: An Analysis of Three
Internet Publications.'' \emph{Linux Gazette} 38: 2.

\hypertarget{ref-copernicoooo}{}
Benito, Juan Cruz. 2017. ``Copernicoooo.'' Universidad de Salamanca.
\url{http://nbviewer.jupyter.org/github/juan-cb/copernicoooo/blob/2c469c6efdf61d7c94a17aba8ea714c76fa1e76d/copernicoooo.ipynb}.

\hypertarget{ref-blanco2009blog}{}
Blanco, Sonia Ruiz. 2009. ``Del Blog Al Microblog: El Devenir Del
Receptor En Generador Y Emisor de Contenidos En La Web 2.0.''
PhD thesis, Universidad de Málaga.

\hypertarget{ref-dans2008anatomia}{}
Dans, E. 2008. ``Anatomía Del Efecto Menéame.'' \emph{Recuperado a
Partir de Http://Www. Enriquedans.
Com/2008/08/Anatomia-Del-Efecto-Meneame. Html}.

\hypertarget{ref-halavais2001slashdot}{}
Halavais, Alexander M Campbell. 2001. ``The Slashdot Effect: Analysis of
a Large-Scale Public Conversation on the World Wide Web.'' PhD thesis.

\hypertarget{ref-Merelo2017:figshare}{}
Merelo, Juan J. 2017a. ``Análisis de la duración de la fama de la página
de Nicolás Copérnico en la Wikipedia española,'' January.
doi:\href{https://doi.org/10.6084/m9.figshare.4535372.v3}{10.6084/m9.figshare.4535372.v3}.

\hypertarget{ref-merelo17}{}
---------. 2017b. ``Copérnico Famoso Por Un Día.'' 1. GeNeura team,
Universidad de Granada.

\hypertarget{ref-merelo17:2}{}
---------. 2017c. ``Evolución de La Notoriedad de La Página de Nicolás
Copérnico En La Wikipedia Española.'' 2. GeNeura team, Universidad de
Granada.

\hypertarget{ref-naaman2011hip}{}
Naaman, Mor, Hila Becker, and Luis Gravano. 2011. ``Hip and Trendy:
Characterizing Emerging Trends on Twitter.'' \emph{Journal of the
Association for Information Science and Technology} 62 (5). Wiley Online
Library: 902--18.

\hypertarget{ref-wiki:Copernico}{}
Wikipedia. 2017. ``Nicolás Copérnico --- Wikipedia, La Enciclopedia
Libre.''
\url{https://es.wikipedia.org/w/index.php?title=Nicol\%C3\%A1s_Cop\%C3\%A9rnico\&oldid=96120597}.

\end{document}